\documentclass[aps,prl,showpacs,noshowkeys,amsmath,amssymb,amsfonts,superscriptaddress,twocolumn]{revtex4-2}

\usepackage{amsmath,amssymb,physics,graphicx,bm,siunitx}
\usepackage{xcolor}
\usepackage[colorlinks=true,linkcolor=blue,citecolor=blue,urlcolor=blue]{hyperref}
\usepackage[all]{hypcap} 
\usepackage{graphicx}
\usepackage{amsmath}
\usepackage{bm}
\usepackage{lipsum}
\usepackage{amsmath}
\usepackage{upgreek}
\usepackage{color}
\usepackage{soul} % for strikeout

\definecolor{darkgreen}{rgb}{0.2,0.6,0.2}

\begin{document}

\title{Engineering tunable fractional Shapiro steps in colloidal transport }

\author{Andris P. Stikuts}
\email{Both authors equally contributed to this work.}
\affiliation{Departament de F\'{i}sica de la Mat\`{e}ria Condensada,
Universitat de Barcelona, 08028, Spain}
\affiliation{University of Barcelona Institute of Complex Systems (UBICS),
08028, Barcelona, Spain}

\author{Seemant Mishra}
\email{Both authors equally contributed to this work.}
\affiliation{Universit\"{a}t Osnabr\"{u}ck, Institut f\"ur Physik,
Barbarastra{\ss}e 7, D-49076 Osnabr\"uck, Germany}

\author{Artem Ryabov}
\email{artem.ryabov@mff.cuni.cz}
\affiliation{Charles University, Faculty of Mathematics and Physics, 
Department of Macromolecular Physics, V Hole\v{s}ovi\v{c}k\'{a}ch 2, 
CZ-18000 Praha 8, Czech Republic}

\author{Philipp Maass}
\email{maass@uos.de}
\affiliation{Universit\"{a}t Osnabr\"{u}ck, Institut f\"ur Physik,
Barbarastra{\ss}e 7, D-49076 Osnabr\"uck, Germany}

\author{Pietro Tierno}
\email{ptierno@ub.edu}
\affiliation{Departament de F\'{i}sica de la Mat\`{e}ria Condensada, Universitat de Barcelona, 08028, Spain}
\affiliation{University of Barcelona Institute of Complex Systems (UBICS), 08028, Barcelona, Spain}

\date{\today}
%\pacs{64.75.Gh, 05.70.Fh, 64.70.K-}

\maketitle \textbf{Shapiro steps are quantized plateaus in the velocity-force or velocity-torque curve of a driven system, when its speed remains constant despite an increase in the driving force. For microscopic particles driven across a sinusoidal potential, integer Shapiro steps have been observed.
By driving a single colloidal particle across a time-modulated, non-sinusoidal periodic optical landscape, we here
demonstrate that fractional Shapiro steps emerge in addition to integer ones.
Measuring the particle position via 
individual particle tracking, we reveal the underlying microscopic mechanisms that produce integer and fractional steps and demonstrate how these steps can be controlled by tuning the shape and driving protocol of the optical potential. 
The flexibility offered by optical engineering allows us to generate wide ranges of potential shapes and to study,
at the single-particle level, synchronization behavior in driven soft
condensed matter systems.}

\vspace{2ex}\noindent

Many physical systems driven out-of-equilibrium by
an external force are characterized in terms of a velocity-frequency or velocity-force 
relation~\cite{Reimann2002,Haggi2009,Matrasulov2014,Reichhardt2016},
which is an analogue of the current-voltage characteristic measured in superconducting junctions~\cite{Rowell1963,Golubov2004}, carbon nanotubes~\cite{Sander1997,Kasumov1999,Balasubramanian2005}, graphene~\cite{Novoselov2004,Heersche2007} and other electronic circuits~\cite{Floyd1994}. 
When subjected to a time-periodic force, the internal dynamics of these systems may synchronize with the external driving. 
The synchronization effect manifests itself in the form of constant plateaus of voltage or velocity versus mean current or force. 
Such plateaus, known as Shapiro steps, were first reported for a superconducting
Josephson junction driven by a microwave
signal~\cite{Shapiro1963,Grimes1968}, where discrete steps in the voltage arise from the synchronization
between the applied signal and oscillations of the Josephson phase~\cite{Waldram1976,Likharev1979}.  

Since their discovery,
understanding the nature of Shapiro steps has been important not
only for elucidating basic mechanisms underlying superconductivity, but also
for new technological developments including, for example, the
realization of metrological voltage
controllers~\cite{Burroughs1999,Burroughs2011}.

In classical driven systems, the presence of discrete steps in the
velocity-force curves
can be the signature of different physical effects, from
speed reduction due to friction~\cite{Braun1998,Vanossi2013,Oded2018},
to the onset of a pinning-depinning transition
\cite{Fisher1998,Brazovskii2007},
synchronization~\cite{Pikovsky2001,Acebron2005} or
locking~\cite{Wiersig2001,Reichhardt2002,Creighton2007} with an
underlying energetic landscape.  
In many-particle systems, 
collective particle motions through
periodic~\cite{Harada1996,Voit2000,Bloch2008,Custer2020} or
random~\cite{Bechinger2016,Reichhardt/etal:2022} landscapes are often mediated
by defects, which move like a single particle and
display a complex sequence of plateaus with integer or
fractional values of their speed.
Due the subtle interplay of interparticle and particle-substrate interaction, 
defect propagation takes place in complex periodic energy landscapes.

Modern
advancements in optical manipulation of microscale matter have made it
possible to engineer such potentials with tunable energetic wells and
inter-well distances~\cite{Grier2003,Kishan2010,Padgett2011}.  With
this capability one can manipulate and drag microscopic
particles~\cite{Ashkin1986,Molloy2002}, biological
systems~\cite{Ashkin1987,Svoboda1993}, measure tiny
forces~\cite{Brunner2004,Pinyu2009}, or even assemble matter in
two~\cite{Baumgartl2007,Mikhael2008} or three
dimensions~\cite{Leach2004,Lee2007,Melzer2021}.  In this context,
integer Shapiro steps have been recently reported for a single
colloidal particle driven through an optical sinusoidal potential by an
underlying oscillating substrate~\cite{Juniper2015}.  The plateaus
emerged due to dynamic mode locking~\cite{Pikovsky2001}, where
synchronization with the oscillating substrate leads to a sequence of
transport modes characterized by a constant speed.
 
By engineering non-sinusoidal periodic potentials with several maxima per wavelength,
we show here that a single particle 
can display integer and fractional steps in the average speed
under time-dependent driving. 
Employing individual particle tracking, we unveil
the microscopic mechanisms leading to 
the emergence of these steps, and
show how it is possible to control them by constructing
diagrams of phase-locked modes.  Our driving strategy implements the
time modulation directly within the periodic potential, without the
need to translate or oscillate the substrate which could induce delay
in the particle response and hydrodynamic back-flow.  The 
non-sinusoidal shape allows the particle to synchronize with the driving potential in a variety
of modes, generating both pronounced integer and fractional Shapiro
steps.  Using our ability to tune the optical landscape and the driving protocol, we can even 
increase the prominence of some fractional Shapiro steps over others.

%%%%%%%%%%%%%%%%%%%%%%%%%
% Fig1
%%%%%%%%%%%%%%%%%%%%%%%%%
\begin{figure*}[t]
\includegraphics[width=\textwidth]{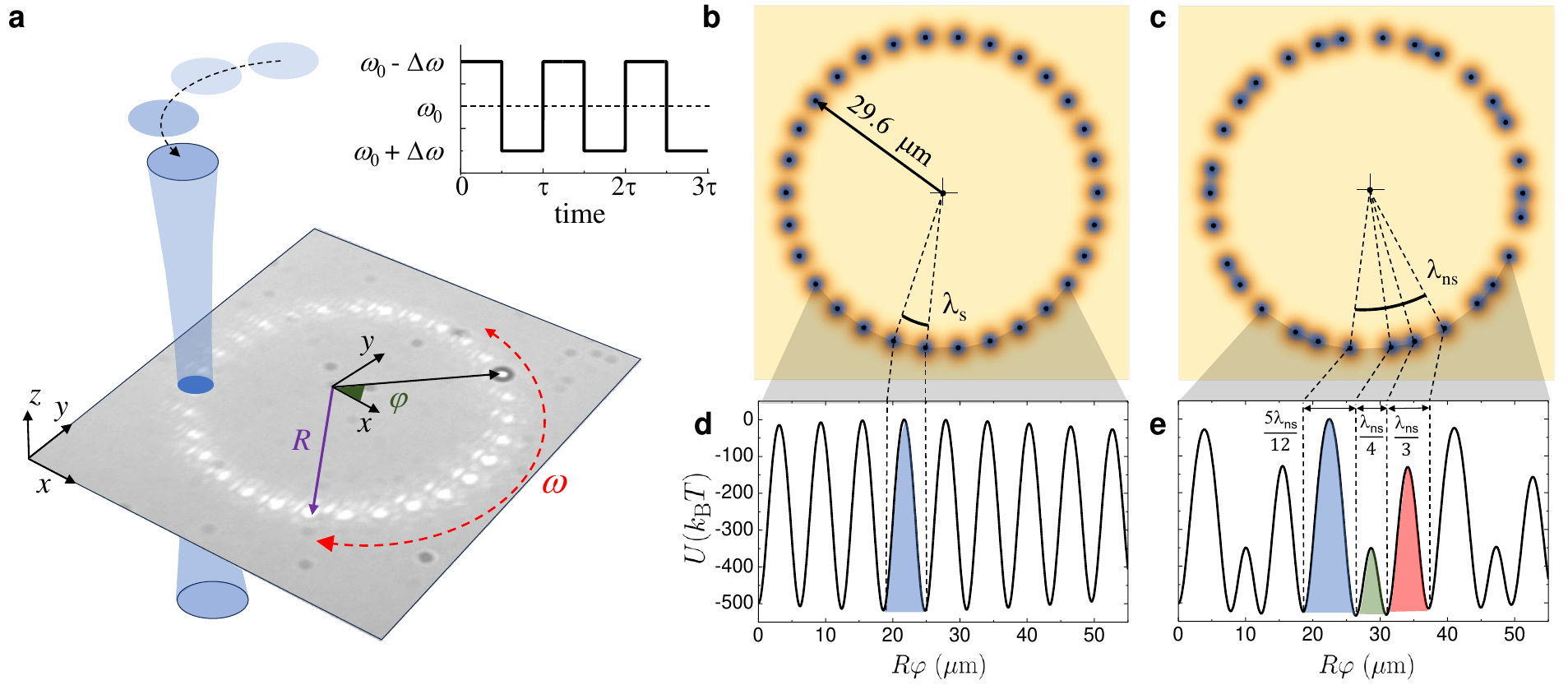}
\caption{\textbf{Time-modulated sinusoidal and
    non-sinusoidal optical potentials.}  \textbf{a}, Colloidal
  particle driven along a rotating ring of optical traps created by
  fast scanning tweezers. The ring rotates with a time-dependent
  angular frequency $\omega(t)$ modulated according to a square wave.
  Inset at the top displays the modulation with period $\tau$ and
  amplitude $\Delta \omega$ around a mean frequency $\omega_0$.
  \textbf{b},\textbf{c}, Density plots of $N_{\rm tr}=30$ optical traps with
  positions on a circle of radius $R =29.6\, \rm{\mu m}$.  In
  \textbf{b}, the optical traps are arranged equidistantly and create a sinusoidal
  potential with wavelength $ \lambda_{\rm s}= 2\pi R/30$ along the
  ring.  In \textbf{c}, the traps are arranged in 10 groups of triplets,
  yielding a periodic non-sinusoidal potential with wavelength $
  \lambda_{\rm ns}= 2\pi R/10$.  In each triplet, spacing between
  neighboring traps are $5\lambda_{\rm ns}/12$, $\lambda_{\rm ns}/4$
  and $\lambda_{\rm ns}/3$.  \textbf{d},\textbf{e}, Corresponding
  profiles of the optical potential  $U$
  extracted from
  torque measurements.  Supplementary Videos 1-6 in the Supporting Information show 
  representative motions of the particle in these two potentials.}
\label{figure1}
\end{figure*}
%%%%%%%%%%%%%%%%%%%%%%%%%

%\section*{Realization of time-modulated\\ periodic potential}
\section*{Designing time-modulated periodic\\ optical potentials}
We
drive a single polystyrene colloidal particle of diameter
$\sigma=4\,\rm{\mu m}$ across a time-dependent periodic potential
generated by passing an infrared continuous-wave laser through a pair
of acousto-optic deflectors, as sketched in Fig.~\ref{figure1}a. 
The particles sediment close to the bottom surface where they 
float due to balance between gravity and electrostatic repulsive interactions. Since the particles are illuminated by the laser from the top, they are trapped in two dimensions, i.e. close to the bottom plane.
Technical details are given in the Methods Section.  

The laser is
rapidly steered through $N_{\rm tr}=30$ positions on a circle of
radius $R=29.6 \, \rm{\mu m}$. Each trap position is visited every $20
\, \rm{\mu  s}$ \cite{Cereceda2021, Cereceda2022,
  Cereceda2023}, a time scale much smaller than the self-diffusion time
 of the particle in the absence of driving, $\sigma^2 / D\sim 267\, {\rm s}$, 
 where $D\sim0.057 \, \rm{\mu m^2 \, s^{-1}}$ is the
self-diffusion coefficient. As the scanning is much faster than the
particle motion, the particle feels an effective, quasistatic
potential.
When a laser passes through a position along the circle, its spot creates a Gaussian potential well with
a depth proportional to the laser power. 
As described in the Methods section,  
the periodic potential felt by the colloidal particle results from the 
superposition of these Gaussian wells.

We engineer two types of periodic potentials, a sinusoidal one $U_{\rm
  s}$ with wavelength $\lambda_{\rm s}=2\pi R/30$,
Fig.~\ref{figure1}b,d, and a non-sinusoidal one $U_{\rm ns}$ with
wavelength $\lambda_{\rm ns}=3\lambda_{\rm s}$, Fig.~\ref{figure1}c,e.
The $U_{\rm ns}$ is generated by placing the traps in groups of
three per wavelength $\lambda_{\rm ns}$, with trap centers located at
positions $0$, $\lambda_{\rm ns}/3$, and $7\lambda_{\rm ns}/12$ in
$[0,\lambda_{\rm ns})$.  With this arrangement, we obtain three distinct potential
    wells per wavelength $\lambda_{\rm ns}$ with inter-well spacings
    $\lambda_{\rm ns}/3$, $\lambda_{\rm ns}/4$, and $5\lambda_{\rm ns}/12$.  
    The
    profiles of the periodic potentials are calculated from
    measured particle trajectories, see Methods Section for 
    details.  These profiles, shown in 
    Fig.~\ref{figure1}d for the
    sinusoidal potential and in Fig.~\ref{figure1}e for the
    non-sinusoidal one, exhibit energetic barriers of about 
    $500\,k_{\rm B}T$. Thus, thermal fluctuations
    are rather weak in our system.

The selected distances between the optical traps were chosen such that the overlap of the Gaussian potential wells generate a simple periodic but non-sinusoidal potential able to induce integer and fractional Shapiro steps in the particle current. The distance between the potential wells where chosen such that it allows to generate two potential wells each of them able to stably trap a colloidal particle. Indeed, distance smaller than the particle diameter could have been not distinguished as different by the particle which would feel them as a single, larger well rather than two distinct ones. On the other hand, potential wells very far from each other could be unable to stably trap the particle along a ring, due to small but still present thermal fluctuations along the radial direction.

By rotating all traps with an angular speed $\omega$, the particle is
driven along the circle, as illustrated in Fig.~\ref{figure1}a.  For
inducing Shapiro steps in the average particle velocity, $\omega$ is
modulated periodically in time with period $\tau$, yielding a
time-modulated driving of the particle.  Specifically, as shown in Fig.~\ref{figure1}a, we apply a
square-wave protocol with mean frequency $\omega_0$ and amplitude
$\Delta\omega$: $\omega(t) = \omega_0+\Delta \omega$ for
$t\in[n\tau,(n+1/2)\tau)$ and $\omega(t) = \omega_0-\Delta \omega$ for
    $t\in[(n+1/2)\tau,(n+1)\tau)$, $n=0,1,2,\ldots$.

%%%%%%%%%%%%%%%%%%%%%%%%%
% Fig2
%%%%%%%%%%%%%%%%%%%%%%%%%
\begin{figure*}[t]
\includegraphics[width =0.8\textwidth]{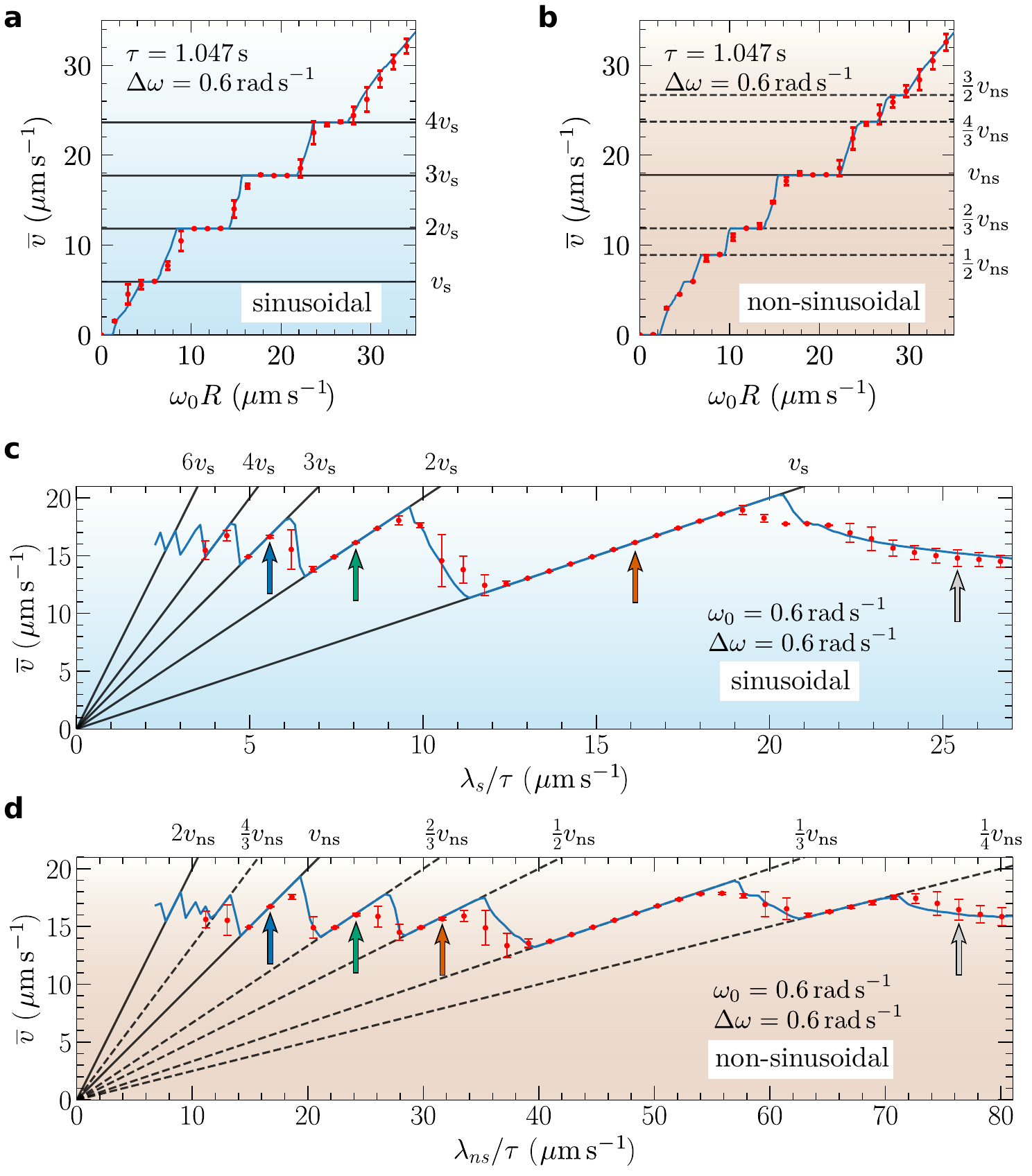}
\caption{\textbf{Phase-locked mean particle velocities reflecting
    motion synchronized with the driving.}  \textbf{a},\textbf{b},
  Average velocity $\bar v$ in the corotating frame versus mean
  azimuthal velocity $\omega_0 R$ of trap rotation for the sinusoidal
  (\textbf{a}) and the periodic but non-sinusoidal optical potential
  (\textbf{b}). In both images we vary $\omega_0$ at fixed 
  $R= 29.6 \rm{\mu m}$, $\tau=1.047\,{\rm s}$, and 
  $\Delta \omega= 0.6 \,{\rm rad\, s}^{-1}$.
  \textbf{c},\textbf{d}, Average velocity $\bar v$ as a function of
  the characteristic oscillatory driving velocities 
  $\lambda_{\rm s}/\tau=v_{\rm s}$ and $\lambda_{\rm ns}/\tau=v_{\rm ns}$ for the
  sinusoidal potential (\textbf{c}) and the non-sinusoidal one
  (\textbf{d}).  
    In both images we vary $\tau$ at  fixed 
    $\omega_0=\Delta\omega_0=0.6\,{\rm
    {rad}\,s}^{-1}$, and $R=29.6\,\rm{\mu m}$.  
     In all graphs scattered red dots are experimental data,
  the error bars denote the standard deviation of three experimental
  measurements, and blue lines are results from numerical
  simulations. Continuous black lines mark integer steps and dashed
  lines mark fractional ones.  Values on top of the graphs in
  \textbf{c},\textbf{d} indicate slopes $n$ and $p/q$ of the 
  lines according to equations~\eqref{eq:v_n} and \eqref{eq:v_pq}, the colored arrows denote measurements, whose particle trajectories are shown in Fig.~\ref{figure3}. }
\label{figure2}
\end{figure*}
%%%%%%%%%%%%%%%%%%%%%%%%%

\section*{Shapiro steps in mean particle velocity} 

For a constant
angular velocity $\omega_0$, the particle trapped in one of the
potential wells tends to follow the trap rotation in clockwise
direction but experiences a resistance due to the Stokesian friction
exerted by the surrounding water. This friction gives rise to a torque
in counter-clockwise direction.  Since this torque is a constant proportional to
$\omega_0$ in a frame corotating with the traps, we analyze the
particle motion in this frame. The average velocity $\bar v$ along
the tangential direction in the corotating frame corresponds to an
average velocity $\bar v_{\rm lab}=\bar v-\omega_0 R$ in the
laboratory frame.

Applying the square-wave modulation $\omega(t)$, we measure $\bar v$
for both the sinusoidal [Fig.~\ref{figure1}d] and the non-sinusoidal
potential [Fig.~\ref{figure1}e].  Figures~\ref{figure2}a,b show
results of these measurements for varying $\omega_0$ at fixed period
$\tau=1.047\,\si{s}$ and amplitude $\Delta \omega= 0.6\,\si{rad\,
  s^{-1}}$ of the driving (red circles with error bars). In both
figures, intervals of $\omega_0$ occur, where $\bar v$ remains
constant corresponding to a sequence of Shapiro steps. In these
plateau regimes of constant $\bar v$, the particle synchronizes its
motion with the oscillatory driving, leading to phase-locked particle
velocities in the corotating reference frame.  The experimental
results are in excellent agreement with Brownian dynamics simulations
(blue lines) detailed in the Methods Section.

For the sinusoidal potential [Fig.~\ref{figure2}a], all steps are
integer multiples of the characteristic speed $v_{\rm s}= \lambda_{\rm
  s}/\tau=5.92\,\si{\mu m\,s^{-1}}$ of the driving, yielding phase-locked
values
\begin{equation}
v_n = n v_{\rm s}, \quad n=1,2,\ldots
\label{eq:v_n}
\end{equation}
These steps are indicated by the horizontal black lines in
Fig.~\ref{figure2}a for $n=1,\ldots4$.  While not all steps are equally well
pronounced, only integer plateaus were observed.
%\blue{The integer steps arise from a synchronized particle motion, where in one period $\tau$ of the driving, the particle is displaced by 
%$n$ wavelengths $\lambda_{\rm s}$ of the sinusoidal potential.}
These steps arise from a synchronized particle motion, where in one period $\tau$ of the driving, the particle is displaced by 
$n$ wavelengths $\lambda_{\rm s}$ of the sinusoidal potential.

%SUGGESTION
For the non-sinusoidal potential [Fig.~\ref{figure2}b], we observe steps (dashed horizontal
black line)
%For the non-sinusoidal potential [Fig.~\ref{figure2}b], in addition to integer steps (solid horizontal
%black line) where $v_{\rm
%  ns}=\lambda_{\rm ns}/\tau=17.76\,\si{\mu m\,s^{-1}}$ (solid horizontal
%black line), we observe fractional steps
\begin{equation}
v_{p,q}= \frac{p}{q}\,v_{\rm ns} \, \, \, .
\label{eq:v_pq}
\end{equation}
which are a fractions $p/q$ of the characteristic velocity $\bar v=v_{\rm
  ns}=\lambda_{\rm ns}/\tau=17.76\,\si{\mu m\,s^{-1}}$. Also, the integer
  step with $\bar v=v_{\rm
  ns}=\lambda_{\rm ns}/\tau$ occurs (solid horizontal
black line).
The fractional plateaus arise from a synchronized particle motion, where in $q$ periods $\tau$ of the driving, the particle is displaced by 
$p$ wavelengths $\lambda_{\rm ns}$ of the non-sinusoidal potential.
%The fractional steps in Fig.~\ref{figure2}b are shorter than integer ones,}
%with $p/q=1/2$, 2/3, 4/3, and 3/2 (dashed horizontal black lines).

The fractional plateaus are smaller than integer ones, and thus more
challenging to be resolved. In our experiments, we increased $\omega_0
R$ in steps of $1.5\,\si{\mu m\, s^{-1}}$. For this resolution, none of
the fractional steps in Fig.~\ref{figure2}b covers an interval of
$\omega_0 R$ consisting of more than two experimental points.
However, the numerical simulations shown by the blue lines in Figs.~\ref{figure2}b provide evidence of 
the presence of the fractional steps in the non-sinusoidal potential in Fig.~\ref{figure2}b.

%\blue{To identify both integer and fractional Shapiro steps in the average particle speed, we first measure  
%$\bar{v}$ by varying the angular velocity $\omega_0$. This representation is similar to the velocity-force or velocity-torque curves reported in several condensed matter systems. 
%SUGGESTION
%\blue{The representation of $\bar{v}$ as a function of $\omega_0 R$ 
%%[or torque ...] 
%in Figs.~\ref{figure2}a,b corresponds
%to the velocity-force curves reported in several condensed matter systems. 
%However, as shown in Fig.~\ref{figure2}b, fractional plateaus are smaller than integer ones in this representation, and more challenging to be resolved. 
%We increased $\omega_0 R$ in steps of $1.5\,\si{\mu m\, s^{-1}}$. With this resolution, none of the fractional steps in Fig.~\ref{figure2}b covers an interval of $\omega_0 R$ consisting of more than two experimental points.  However, the numerical simulations shown by the blue lines in Figs.~\ref{figure2}b provide evidence of the presence of the fractional steps in the non-sinusoidal potential in Fig.~\ref{figure2}b.}

Nonetheless, to unambiguously demonstrate synchronized motion with
phase-locked velocities according to equations~\eqref{eq:v_n} and
\eqref{eq:v_pq} in our experiments, we can also keep $\omega_0$ and
$\Delta\omega$ fixed and vary the period $\tau$ of the driving.  For a
good choice of $\omega_0$ and $\Delta\omega$, the driving should be
such that the particle can surmount the potential barriers at the
square wave's high value $\omega_0+\Delta\omega$, while it remains
trapped in a potential well at the low value $\omega_0-\Delta\omega$.
This can be ensured for the trapping by taking
$\Delta\omega=\omega_0$.  For surmounting barriers, we choose
$\omega_0=0.6\,{\rm rad\,s}^{-1}$, which is comparable to the critical
frequency at which the particle starts slipping in the potential.

Figures~\ref{figure2}c,d show the mean particle velocity as a function
of the characteristic velocities $\lambda_{\rm s}/\tau$ and
$\lambda_{\rm ns}/\tau$ for $\omega_0=\Delta \omega=0.6\,{\rm
  rad\,s}^{-1}$. 
Now, phase-locking according to
Eqs.~\eqref{eq:v_n} and \eqref{eq:v_pq}
manifests itself as
$\bar v$ varying linearly with $v_{\rm s}=\lambda_{\rm s}/\tau$ and  $v_{\rm ns}=\lambda_{\rm ns}/\tau$.
Fractional phase locking in Fig.~\ref{figure2}d is
much more clearly visible than in Fig.~\ref{figure2}b:
a linear increase of $\bar v$ with $\lambda_{\rm ns}/\tau$
occurs in broad intervals with fractional slopes $p/q=1/4$, $1/3$, $1/2$,
and $2/3$.
Again, simulations
(blue lines) in both Figs.~\ref{figure2}c,d are in excellent agreement
with the experimental observations.

The error bars in Figs.~\ref{figure2}a-d  represent the spread between three different sets of measurements, each conducted 
with a different particle from the same stock solution. 
They are relatively small when the particle is 
phase-locked with the oscillating potential and become larger when it is not. 
This can be understood from the fact that for synchronized motion, fluctuations of the particle position 
due to thermal noise are suppressed. The effect is reflected in a lower diffusion coefficient 
of the particle in a phase-locked state \cite{Juniper2017}.

%%%%%%%%%%%%%%%%%%%%%%%%%
% Fig3
%%%%%%%%%%%%%%%%%%%%%%%%%
\begin{figure*}[t]
\includegraphics[width=\textwidth]{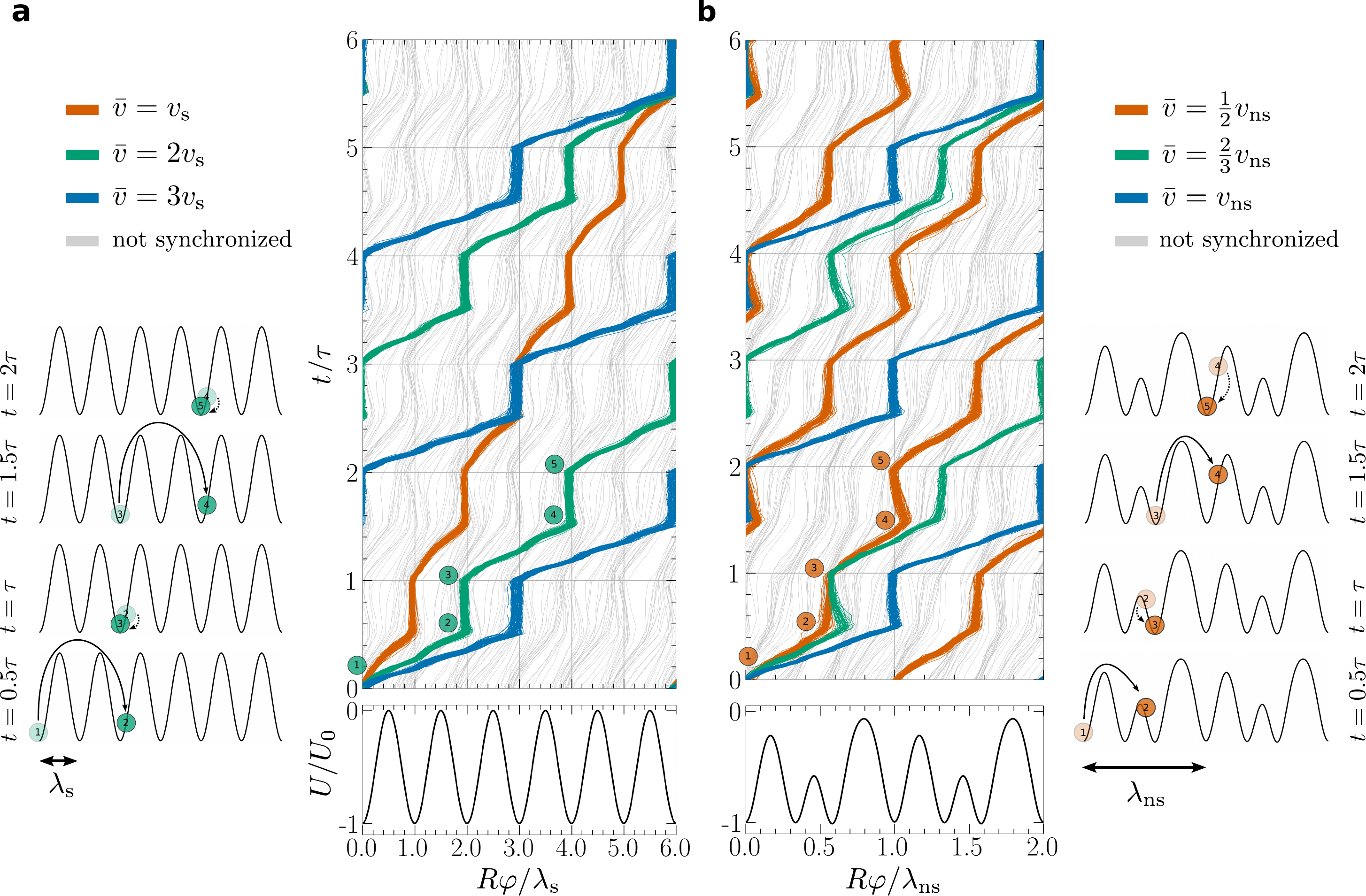}
\caption{\textbf{Synchronized and non-synchronized particle
    trajectories.}  \textbf{a},\textbf{b}, 
Scaled angular particle
  positions $R\varphi /\lambda_{\rm s,ns}$ versus scaled time $t/\tau$
  in the corotating reference frame at fixed $\omega_0=\Delta \omega = 0.60\,\si{rad\,s^{-1}}$
  as in Figs.~\ref{figure2}c,d for (\textbf{a}) the sinusoidal 
  and (\textbf{b}) the non-sinusoidal potential for 
   various driving frequencies $1/\tau$. 
The chosen frequencies correspond to values of $\lambda_{\rm s}/\tau$ and $\lambda_{\rm ns}/\tau$ indicated by the colored arrows in Figs.~\ref{figure2}c,d.
  Optical potentials are shown at the bottom. 
  Trajectories are displayed in a reduced zone scheme, i.e.\ when they 
  exit at the right side of the graph they are continued at the left side.
  Synchronized trajectories corresponding to Shapiro steps overlap and
  are colored in orange,
  green and blue. Non-synchronized trajectories are marked in gray.
   In \textbf{a}, mean velocities $\bar v$ obtained from
  the synchronized trajectories are integer multiples of $v_{\rm
    s}$. In \textbf{b}, fractions $p/q$ of $v_{\rm ns}$ appear,
  which correspond to particle displacements by $p$ wavelengths
  $\lambda_{\rm ns}$ in $q$ periods of the driving.
    The small schematics on the side of the images illustrate the particle position 
    within the periodic potential corresponding to an integer step (side of  \textbf{a}) and to a fractional one (side of  \textbf{b}).  
    The videos 1-6 in the Supplementary Information show the synchronized particle motion across the periodic potentials.}
\label{figure3}
\end{figure*}
%%%%%%%%%%%%%%%%%%%%%%

The ability to precisely track the colloidal particle allows us to analyze in detail its movement 
across the two types of periodic potentials. 
In Fig.~\ref{figure3} we show particle trajectories for various driving frequencies $1/\tau$ at fixed 
$\omega_0=\Delta\omega$ for the sinusoidal (Fig.~\ref{figure3}a) and non-sinusoidal potential (Fig.~\ref{figure3}b).

Colored synchronized trajectories collapse onto space-time periodic limit cycles apart from thermal fluctuations. 
Along the synchronized trajectories in Fig.~\ref{figure3}a,
the particle moves $n=1,2,3$ wavelengths $\lambda_{\rm s}$ in one period $\tau$ of the driving, giving a mean particle velocity $\bar v=n v_{\rm s}$.
Along the synchronized trajectories in Fig.~\ref{figure3}b,
the particle moves $p$ wavelengths $\lambda_{\rm ns}$ in $q$ periods $\tau$, giving the mean particle velocity $\bar v=(p/q) v_{\rm ns}$ with fractions
$p/q=1/2$, $2/3$, $1/1$. Grey lines in both graphs correspond to particle trajectories that are not synchronized with the driving. 
In the schematics on the sides of Fig.~\ref{figure3}, we illustrate the particle displacements in successive
half periods $\tau/2$ for the phase-locked modes with $n=2$ in Fig.~\ref{figure3}a and $p/q=2/3$ in Fig.~\ref{figure3}b.

For the sinusoidal potential, 
we see that integer
Shapiro steps with $\bar v=v_{\rm s} ,2v_{\rm s}, \ldots$ occur when
the particle is trapped in a potential minimum for half a period of
the driving, and then moves 
a distance $\lambda_{\rm s}, 2\lambda_{\rm s}, \ldots$ in the other half.  
The trapping at the minimum aids the
synchronization as a way of ``resetting'' the particle position during
each period.

An analogous situation occurs for the non-sinusoidal potential,
but now the particle displays an additional backward or forward movement to reach the closest potential minimum.
These additional movements can be seen, for
example, in the bunches of orange particle trajectories 
in Fig.~\ref{figure3}b, which correspond to the step at $\bar
v=\frac{1}{2}v_{\rm ns}$ in Fig.~\ref{figure2}b.

%%%%%%%%%%%%%%%%%%%%%%%%%
% Fig4
%%%%%%%%%%%%%%%%%%%%%%%%%
\begin{figure}[t]
\includegraphics[width=\columnwidth]{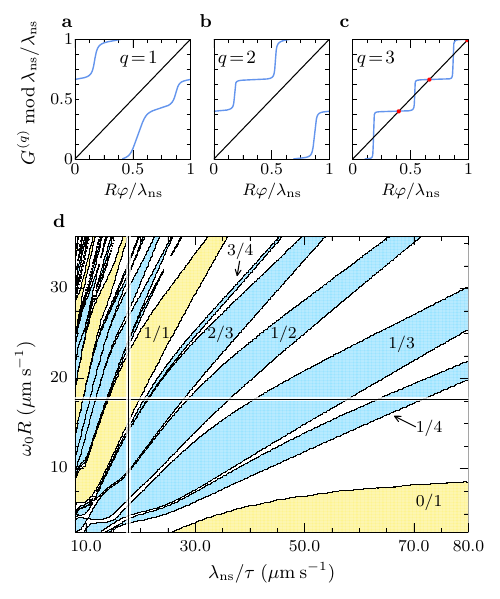}
\caption{\textbf{Diagram of phase-locked modes.} \textbf{a-c},
  Propagators $G^{(q)}(x)$ giving position of the particle after $q$
  periods of the driving if it started at $x$.
  Parameters are as in
  Fig.~\ref{figure2}d when $\lambda_{\rm ns}/\tau=50\,\mu {\rm
    m\,s}^{-1}$.  Intersections of $G^{(3)}(x)$ with the diagonal line in \textbf{c}
  imply that a limit cycle of three stable fixed points (bullets) is
  present, giving $p=|G^{(q)}(x_\ast)-x_\ast|/\lambda_{\rm ns}=1$.
  The theoretically predicted mode of phase locking thus is 1/3, in
  agreement with the experimental finding in Fig.~\ref{figure2}d.
  \textbf{d}, Diagram of phase-locked modes for the non-sinusoidal
  potential obtained by the fixed-point analysis when applying it to a
  wide range of values $\lambda_{\rm ns}/\tau$ and $\omega_0 R$ at
  fixed $\Delta\omega=0.6\,{\rm rad\, s}^{-1}$.  Yellow regions indicate
 integer Shapiro steps and blue regions mark fractional steps. Following the
  vertical and horizontal lines in the diagram corresponds to the
  variations of $\omega_0 R$ and $\lambda_{\rm ns}/\tau$ in
  Figs.~\ref{figure2}b and \ref{figure2}d, respectively.}
\label{figure4}
\end{figure}
%%%%%%%%%%%%%%%%%%%%%%%%%

Next we show that it is possible to calculate the type of phase-locked
modes characterized by $p/q$ in equation~\eqref{eq:v_pq} and thus to
control the experimental parameters that give rise to a selected step.  

We consider the particle
motion in absence of thermal fluctuations and determine the propagator
$G(x)$, which gives the position of the particle after one period of
the driving if it started at position $x=R\varphi$.
After $q'$ periods of the driving, the particle position is obtained
by the $q'$-fold composition $G^{(q')}(x)=G\circ\ldots\circ G(x)$ of
$G(x)$.  For a mode to occur 
with value $q'$, the particle must be at
an equivalent position in the potential after $q'$ periods of
the driving.
This means that there must exist an
$x_\ast\in[0,\lambda_{\rm ns})$, where the difference
    $G^{(q')}(x_\ast)-x_\ast$ is an integer multiple of $\lambda_{\rm
      ns}$, i.e.\ such $x_\ast$ must satisfy the fixed point equation
\begin{equation}
G^{(q')}(x_\ast)\hspace{-0.7em}\mod \lambda_{\rm ns}=x_\ast\,,
\label{eq:Gq-fixed-points}
\end{equation}
where $a\hspace{-0.5em}\mod b\in[0,b)$ is the remainder when $a$ is
    divided by $b$.  The smallest $q'$, for which at least one stable
    fixed point $x_\ast$ exists, is the $q$ of a phase-locked mode.
    The number $p$ of wavelengths $\lambda_{\rm ns}$ by which the
    particle is displaced after $q$ periods is
    $p=|G^{(q)}(x_\ast)-x_\ast|/\lambda_{\rm ns}$.  This fixed-point
    method allows us to calculate $q$ and $p$ for any given driving
    parameters $\omega_0$, $\Delta\omega$, and $\tau$.  Technical
    details are given in the Methods Section.
 
Figures~\ref{figure4}a-c demonstrate the application of the
fixed-point method for the non-sinusoidal potential and the same 
parameters as in Fig.~\ref{figure2}d when $\lambda_{\rm
  ns}/\tau=50\,\mu {\rm m\,s}^{-1}$.  For $q=1$ [Fig.~\ref{figure4}a]
and $q=2$ [Fig.~\ref{figure4}b], no fixed point solutions of
equation~\eqref{eq:Gq-fixed-points} exist, while a limit cycle of
three stable fixed points $x_\ast$ (bullets) is obtained for $q=3$ 
[Fig.~\ref{figure4}c],
giving $p=|G^{(q)}(x_\ast)-x_\ast|/\lambda_{\rm ns}=1$. Accordingly,
the theoretically predicted mode is 1/3, in agreement with the
experimental observation.

Figure~\ref{figure4}d shows the diagram of phase-locked modes when
varying $\omega_0 R$ and $\lambda_{\rm ns}\tau$, and setting
$\Delta\omega=0.6\,{\rm rad\, s}^{-1}$.
The vertical and horizontal lines in this
diagram represent the variation of parameters considered in
Figs.~\ref{figure2}b,d and show good agreement with the experimental
observations: for example, the modes with $p/q=1/2$, 2/3, and 1 at
fixed $\tau=1.047\,{\rm s}$ are predicted to occur for $\omega_0R$ in
the intervals $7-10\,\si{\mu m\, s^{-1}}$, $10-14\,\si{\mu m\,
  s^{-1}}$, and $15-23\,\si{\mu m\, s^{-1}}$ according to
Fig.~\ref{figure4}d, which match the intervals where the modes occur
in Fig.~\ref{figure2}b. Likewise, the modes with $p/q=2/3$, 1/2, and
1/3 at fixed $\omega_0 R=17.76\,\si{\mu m\, s^{-1}}$ are predicted to
occur for $\lambda_{\rm ns}/\tau$ in the intervals $20-28\,\si{\mu m\,
  s^{-1}}$, $29-36\,\si{\mu m\, s^{-1}}$ and $42-57\,\si{\mu m\,
  s^{-1}}$, matching the intervals in Fig.~\ref{figure2}d.

\section*{Conclusions}

We report the observation of integer and fractional Shapiro steps in the 
average speed of a single colloidal particle driven across a spatially and temporally modulated  potential landscape. 
Fractional steps appear only when the potential is not sinusoidal. 
Through direct measurement of the particle and trap positions, and using theoretical arguments, 
we unveil the phase-locking mechanisms at the origin of the fractional steps and demonstrate
the possibility to tune them by engineering the optical driving.

In our periodic potentials, Shapiro steps emerge
via the following mechanisms.
During the part of the period when the driving is zero, the optical potential has a stabilizing role for both integer and fractional steps.
For the sinusoidal potential, the particle relaxes once per driving period, giving rise to integer steps. 
For the non-sinusoidal potential,
due to the complexity of the potential landscape, the particle can relax towards different potential minima in
each driving period. This leads to a richer particle dynamics, enabling the synchronization 
in fractional Shapiro steps in addition to integer ones.

From the application point of view, we have shown how to manipulate the lengths of the fractional plateaus via potential engineering, which is important for particle transport in materials and devices. The occurrence of fractional steps in periodic potentials with anharmonicities is generic. 
They can be used for a versatile steering of stable particle velocities that are robust against noise as well as small perturbations of driving parameters and spatial periodicity. This should allow, for example, for setting a prescribed mean velocity of a particle driven by an external signal within a microfluidic or a lab on a chip-device. 
Another potential application is to use fractional steps for probing characteristic features of periodic potentials, like its deviation from a sinusoidal form, symmetry, or number of wells per period. As diagrams of phase-locked modes are very sensitive to the shape of the potential, the corresponding measurements can 
be employed as a way to determine carefully the parameters in periodic force fields. 
Thus, these modes can be used in sensor applications to detect subtle shifts in the external force, in a similar way than metrological voltage controllers developed for Josephson junctions~\cite{Burroughs1999,Burroughs2011}.

%We expect fractional synchronization modes to be an integral part also in the description of 
%collective excitations in highly dense periodic or quasi-periodic environments 
%by quasiparticles under time-periodic driving.}

%\blue{From the application point of view, we have shown how to manipulate the lengths of the fractional plateaus via potential engineering, which is important for particle transport in materials and devices. Indeed, both integer and fractional plateaus are signatures of synchronized motion with an underlying potential, but fractional one emerges only when such potential is non-sinusoidal. Sinusoidal potentials only describe generically realistic systems under strong approximations. A more common situation would consider an effective periodic but non-sinusoidal potential emerging for example, from the presence of more harmonics which comes from variation of a crystalline structure of a lattice due to disorder. In contrast to integer plateaus, fractional ones provide a more precise control of the synchronization modes and the possibility to switch between different modes which are robust against small changes of the driving parameters. This would allow for setting a prescribed mean velocity of a particle driven by an external signal within a microfluidic or a lab on a chip-device. Also, fractional plateaus detected from the particle trajectory could enable precise measurement of an external periodic force, a mechanical vibration or an electromagnetic signal. }

Unraveling microscopic synchronization mechanisms leading to Shapiro steps 
is important in the analysis of transport 
of particles across periodically structured landscapes, a generic situation encountered
when studying non-equilibrium dynamics in condensed matter systems.
Plateaus are not just a signature of non-linearity; they offer a window for exploring details of the coupling between the particle's motion and
that of the underlying landscape.

%\blue{Fractional Shapiro steps have been experimentally reported in the past mostly in quantum mechanical systems such as superconducting Josephson junctions or charge density waves, but not in classical ones.  In contrast to integer steps, fractional ones allow to identify new synchronization regimes and exotic dynamical states that integer do not reveal.}

While our work has centered on a single driven particle, future directions may explore complex behavior in
collective motions of many particles,
or across disordered landscapes, which can be easily realized with optical engineering. 
This will further enrich the study of resonant transport and phase locking in periodically driven out-of-equilibrium systems.

\section{References}

\newpage
\clearpage

\noindent\textbf{\large Methods}\\
%\section*{Methods}
\noindent\textbf{Experimental setup.}  We use monodisperse spherical
polystyrene particles with diameter $\sigma=4\,\rm{\mu m}$ (CML,
Molecular Probes).  The particles are dispersed in highly deionized
water (Milli-Q water) at room temperature $T=293$ K, and the suspension is
confined within a fluidic cell assembled with two coverslips separated
by $\sim 100 \, \rm{\mu m}$.  The cell is placed on the stage of a
custom-built optical microscope and exposed to a set of fast scanning
optical tweezers.  The tweezers are created by passing an infrared
laser beam with wavelength $1064 \, \rm{\mu m}$ and power $P= 3\,{\rm
  W}$ (manlight ML5-CW-P/TKS-OTS) through a pair of acousto-optic
deflectors (AODs, AA Optoelectronics DTSXY-400-1064). Combined with
the AODs is a two-channel radio frequency wave generator
(DDSPA2X-D431b-34), which is addressed by a digital output card
(National Instruments cDAQ NI-9403) with a refresh frequency of $150$
kHz.  A Nikon $40\times$ microscope objective (plan Apo), illuminated
by a light emitting diode, and a complementary metal oxide camera
(Ximea MQ003MG-CM) are used to record the particle positions at
$30\,{\rm Hz}$.
\vspace{5px}

\noindent\textbf{Optical potential.}  We define by
$\varphi_i=\phi_i(t)-\theta(t)$, $i=1,\ldots,N_{\rm tr}$, the fixed
azimuthal positions of the $N_{\rm tr}$ trap centers in the co-moving
frame, where $\phi_i(t)$ are the rotating positions in the laboratory
frame and 
\begin{equation}
\theta(t) = -\int_0^t\dd t' \omega(t') \, \, \, .
\end{equation}

Each laser spot creates a Gaussian potential well with a depth $A$
proportional to the laser power, and a width $w$. The total potential
felt by the particle at position $\varphi$ in the corotating frame is
given by the superposition of the $N_{\rm tr}$ Gaussian potential
wells,
\begin{equation}
U_0(\varphi) = -A \sum_{i=1}^{N_{\rm tr}} \exp\left(
  -\frac{R^2(\varphi -\varphi_i)^2}{2w^2} \right)\,.
\label{eq:Uopt1}
\end{equation}

Due to slight imperfection in the optical tweezer setup, the well depth is not perfectly uniform across the ring of traps.
It is weakly modulated with two nearly equidistant local maxima along the
ring, giving rise to an amplitude modulation $\propto\cos(2\varphi)$
of the optical potential.
Taking this weak modulation into account,
the potential in the corotating frame becomes
\begin{equation}
U(\varphi,t) = \bigl(1+\xi \cos[2(\varphi+\theta(t)]\bigr)
U_0(\varphi),
\label{eq:Uopt2}
\end{equation}
where $\xi$ is the strength of the amplitude modulation.  With phase
shifts $\alpha$ and $\beta$ of the amplitude modulation and trap
positions in the fixed laboratory frame, the functional form of the
optical potential is
\begin{equation}
U(\varphi,t)= \bigl(1+\xi \cos[2(\varphi + \theta(t) + \alpha)] \bigr)
U_0(\varphi+\beta).
\label{eq:Uopt3}
\end{equation}

To determine the parameters $\xi$, $\alpha$ and $\beta$, and the
parameters $A$ and $w$ entering $U_0(\varphi+\beta)$ via
equation~\eqref{eq:Uopt1}, we follow the procedure described in
Ref.~\cite{Cereceda2022}.  We rotate the potential landscape with a
constant angular frequency $\omega_0=0.6\,\rm{rad\, s}^{-1}$ large
enough to allow for the particle to cross potential barriers
($\Delta\omega=0$).  The time series of particle's positions $\phi(t)$
in the laboratory frame is recorded for 20~minutes at $30$ frames per
second, that is with a time step of $\delta t=1/30\,{\rm s}$.  With the
positions $\varphi(t)=\phi(t)+\omega_0 t $ in the corotating frame,
the angular velocities $[\varphi (t+\delta t)-\varphi(t)]/\delta t$
are calculated and we extract the torques
\begin{equation}
M(t)=\frac{k_{\rm B} T R^2}{D}\left[ \frac{\varphi ( t+\delta t) -
    \varphi( t)}{\delta t} + \omega_0\right]
\end{equation}
acting on the particle in the cororating frame \cite{Cereceda2022}.

Mean values of torque vary with the position along the ring and are
periodic in time with period $2\pi/\omega_0$.  To determine the mean
torques from the times series $M(t)$, the intervals $[0,2\pi)$ and
     $[0, 2\pi/\omega_0)$ of azimuthal positions and times are divided
        into $N_\varphi=210$ and $N_t=10$ equally sized
        bins. Averaging the $M(t)$ in each bin, we obtain the mean
        torques $\bar M(\varphi,t)$, which must agree with the
        derivative of $U(\varphi,t)$ with respect to $\varphi$,
\begin{equation}\label{eq:torque_experiments}
-\frac{\partial U(\varphi,t)}{ \partial\varphi } = \bar
M(\varphi,t)\,.
\end{equation}
The parameters $A$, $w$, $\xi$, $\alpha$, and $\beta$ are obtained by
fitting $\partial U(\varphi,t)/\partial\varphi$ to the measured $\bar
M(\varphi,t)$ with the least square method. For the sinusoidal potential,
$A=559\,k_{\rm B}T=1.36\,{\rm MJ/mol}$, $w=1.45\,\mu{\rm m}$,
$\xi=0.09$, $\alpha=1.81$, and $\beta=0.23$. For the non-sinusoidal potential,
$A=548\,k_{\rm B}T=1.34\,{\rm MJ/mol}$, $w=1.60\,\mu{\rm m}$,
$\xi=0.10$, $\alpha=0.80$, and $\beta=0.18$.

We checked that our simulation results for the average particle velocities in Fig.~\ref{figure2}
are almost unaffected by
small perturbations of the positions $\varphi_i$ of the optical trap centers  as well as small $i$-dependent 
random modulations of $A$ in Eq.~\eqref{eq:Uopt1}.
\vspace{5px}

\noindent\textbf{Brownian dynamics simulations.}  
Equation~(6) gives the potential $U(\varphi,t)$ in the corotating frame. In the laboratory frame it is $U_{\rm lab}(\phi,t)=U(\phi-\theta(t),t)$,
and the Langevin equation for the overdamped Brownian motion
of the particle reads
\begin{equation}
\gamma R\frac{\dd\phi}{\dd t}
=-\frac{1}{R}\,\frac{\partial U_{\rm lab}(\phi,t)}{\partial\phi}+\sqrt{2k_{\rm B} T\gamma}\,\eta(t)\,.
\label{eq:dvarphi-lab}
\end{equation}
Here, $R\phi$ is the particle displacement along the ring,
$\gamma$ is the Stokesian friction coefficient, and $\eta(t)$ is a thermal noise, modeled by Gaussian white noise process with zero mean
and correlation function $\langle\eta(t)\eta(t')\rangle=\delta(t-t')$.

In the corotating frame with angle variable $\varphi=\phi-\theta(t)$, $\dot\varphi(t)=\dot \phi(t)-\omega(t)$, and
Eq.~\eqref{eq:dvarphi-lab} becomes
\begin{equation}
\frac{\dd\varphi}{\dd t} = -\frac{D}{k_{\rm B}TR^2} \frac{\partial
  U(\varphi,t)}{\partial\varphi} +\omega(t) +\frac{\sqrt{2D}}{R}
\eta(t),
\label{eq:eq_of_motion}
\end{equation}
where $D=k_{\rm B}T/\gamma$ is the diffusion coefficient according to the fluctuation-dissipation theorem.
Equation~\eqref{eq:eq_of_motion} is solved numerically using the
Euler-Maruyama method.

\vspace{5px}

\noindent\textbf{Theoretical prediction of phase-locked modes.}  The
fixed point method can be applied to both the sinusoidal and
non-sinusoidal potential. It relies on the propagator
$G(x)=G^{(1)}(x)$, which is determined by numerical solution of the
Langevin equation~\eqref{eq:eq_of_motion} in the limit of zero noise
[$\eta(t)=0$]. The fixed points of $[G^{(q)}(x)\hspace{-0.4em}\mod
  \lambda]$ given by equation~\eqref{eq:Gq-fixed-points}, with
$\lambda=\lambda_{\rm s}$ or $\lambda=\lambda_{\rm ns}$, are zeros of
the function $H^{(q)}(x)=[G^{(q)}(x)\hspace{-0.4em}\mod \lambda]-x$.

To obtain the zeros of $H^{(q)}(x)$, we divide the interval
$[0,\lambda)$ into $M=50$ equidistant points $x_1<\ldots<x_M$, and
    search for all pairs of successive points $x_i$ and $x_{i+1}$,
    where $H^{(q)}(x_i)$ and $H^{(q)}(x_{i+1})$ have different signs.
    Let $m$ be the number of such pairs $x_i^{(\alpha)}$,
    $x_{i+1}^{(\alpha)}$, $\alpha=1,\ldots, m$.  In each interval
    $[x_i^{(\alpha)}, x_{i+1}^{(\alpha)})$, we determine the point
        $x_0^{(\alpha)}$ of sign change with high accuracy by applying
        the Wijngaarden–Dekker–Brent method
        \cite{Press/etal:2007}. The sign change at $x_0^{(\alpha)}$
        does not necessarily imply that $x_0^{(\alpha)}$ is a zero,
        because $H^{(q)}(x)$ can jump at $x_0^{(\alpha)}$. A zero is
        considered to be present at $x_0^{(\alpha)}$, if
        $|H^{(q)}(x_0^{(\alpha)}-\delta)-H^{(q)}(x_0^{(\alpha)}+\delta)|\le\epsilon$
        for $\delta=10^{-6}\lambda$ and $\epsilon=10^{-3}\lambda$.

For a zero $x_0$ of $H^{(q)}(x)$ to be a stable fixed point $x_\ast$
of $[G^{(q)}(x)\hspace{-0.4em}\mod \lambda]$, it must hold
$|\partial_xG^{(q)}(x_\ast)|<1$ \cite{Strogatz:2015}.  Such stable
fixed point $x_\ast$ corresponds to a limit cycle of
$[G(x)\hspace{-0.4em}\mod \lambda]$ running through $q$ points, which
forms an attractor of the stationary particle motion.  Each of the $q$
points of the limit cycle is a stable fixed point of
$[G^{(q)}(x)\hspace{-0.4em}\mod \lambda]$.  For the example shown in
Fig.~\ref{figure4}c, the three points marked by the bullets are stable
fixed points of one limit cycle.  The two other points, where
$G^{(q)}(x)\hspace{-0.4em}\mod \lambda_{\rm ns}$ intersects with the
diagonal line, are unstable fixed points.

The $q$ of the phase-locked mode is the smallest $q'$, where
$[G^{(q')}(x)\hspace{-0.4em}\mod \lambda]$ exhibits a stable fixed
point. We thus obtain $q$ by starting with $q'=1$ and incrementing it
by one until a stable fixed point occurs, i.e.\ a zero $x_\ast$ of
$H^{(q)}(x)$ satisfying $|\partial_xG^{(q)}(x_\ast)|<1$.

For the mode diagram shown in Fig.~\ref{figure4}d, we have carried out the
analysis for a wide range of parameters $\omega_0 R$, $\lambda_{\rm
  ns}/\tau$, and $q$ values up to four. Parameter regions where either
synchronized motion with $q>4$ or non-synchronized motion occurs, are
marked in white.

\vspace{2ex}
\noindent\textbf{\large Data availability}\\
%\section*{Data availability}
The authors declare that all data supporting the findings of this
study are available within the paper and its Supplementary Information
files or available from the corresponding authors upon request.

\vspace{2ex}
\noindent\textbf{\large Code availability}\\
%\section*{Data availability}
The codes used in this study are available from the corresponding
author upon request.

\vspace{2ex}
\noindent\textbf{\large Acknowledgments}\\
%\section*{Acknowledgments}
This project has received funding from the European Research Council
(ERC) under the European Union's Horizon 2020 research and innovation
programme (grant agreement no.\ 811234).  P.T.\ acknowledge support
the Generalitat de Catalunya under Program ``ICREA Acad\`emia'' and
from the project 2021 SGR 00450.  A.R.\ gratefully acknowledges
financial support by the Czech Science Foundation (Project
No.\ 23-09074L), and S.M., A.R.\ and P.M.\ from the Deutsche
Forschungsgemeinschaft (Project No.\ 521001072). S.M.\ and
P.M.\ further acknowledge the use of a high-performance computing
cluster funded by the Deutsche Forschungsgemeinschaft (Project
No.\ 456666331).

\vspace{2ex}
\noindent\textbf{\large Author Contributions}\\
%\section*{Author Contributions}  
A.S.\ performed the experiments. S.M.\ run the numerical
simulations. A.R., P.M.\ and P.T.\ supervised the work. All authors
discussed the results and commented on the manuscript at all stages.

\vspace{2ex}
\noindent\textbf{\large Competing interests}\\
%\section*{Competing interests}  
The authors declare no competing interests.

\vspace{2ex}
\noindent\textbf{\large Additional information}\\
%\section*{Additional information}  
%\noindent
\textbf{Supplementary Information} is available in the online version
of the paper.

\vspace{1ex}\noindent
\textbf{Correspondence} and requests for materials
related to the experiments should be addressed to P.T.\
(ptierno@ub.edu), related to simulations to
A.R.\ (artem.ryabov@mff.cuni.cz) and P.M.\ (maass@uos.de).

\end{document}